\documentclass[iop]{emulateapj}

\usepackage{graphicx}
\usepackage{float} 
\usepackage{amsmath}
\renewcommand{\d}[0]{\mathbf{d}}
\newcommand{\n}[0]{\mathbf{n}}
\newcommand{\s}[0]{\mathbf{s}}

\newcommand{\N}[0]{\mathbf{N}}

\renewcommand{\S}[0]{\mathbf{S}}

\begin{document}

\title{CMB likelihood approximation for banded probability distributions}

\author{E. Gjerl{\o}w\altaffilmark{1}, K. Mikkelsen\altaffilmark{1},
  H. K. Eriksen\altaffilmark{1}, K. M. G{\'o}rski\altaffilmark{2,3},
  G. Huey\altaffilmark{2}, \\ J. B. Jewell\altaffilmark{2},
  S. K. N{\ae}ss\altaffilmark{1,4}, G. Rocha\altaffilmark{2,5}, 
  D. S. Seljebotn\altaffilmark{1}, I. K. Wehus\altaffilmark{2}}
%\author{E. Gjerl{\o}w\altaffilmark{1} et al.}

\email{eirik.gjerlow@astro.uio.no}

\altaffiltext{1}{Institute of Theoretical Astrophysics, University of
  Oslo, P.O.\ Box 1029 Blindern, N-0315 Oslo, Norway}

\altaffiltext{2}{Jet Propulsion Laboratory, California Institute of
  Technology, 4800 Oak Grove Drive, Pasadena CA 91109, USA}
\altaffiltext{3}{Warsaw University Observatory, Aleje Ujazdowskie
4, 00-478 Warszawa, Poland}

\altaffiltext{4}{Department of Astrophysics, University of Oxford, Keble Road, Oxford OX1 3RH, UK}

\altaffiltext{5}{California Institute of Technology, Pasadena, California, USA}
%\date{Received - / Accepted -}

\begin{abstract}
  We investigate sets of random variables that can be arranged
  sequentially such that a given variable only depends conditionally
  on its immediate predecessor. For such sets, we show that the full
  joint probability distribution may be expressed exclusively in terms
  of uni- and bivariate marginals. Under the assumption that the CMB
  power spectrum likelihood only exhibits correlations within a banded
  multipole range, $\Delta\ell$, we apply this expression to two
  outstanding problems in CMB likelihood analysis. First, we derive a
  statistically well-defined hybrid likelihood estimator, merging two
  independent (e.g., low- and high-$\ell$) likelihoods into a single
  expression that properly accounts for correlations between the
  two. Applying this expression to the WMAP likelihood, we verify that
  the effect of correlations on cosmological parameters in the
  transition region is negligible in terms of cosmological parameters
  for WMAP; the largest relative shift seen for any parameter is
  $0.06\sigma$. However, because this may not hold for other
  experimental setups (e.g., for different instrumental noise
  properties or analysis masks), but must rather be verified on a
  case-by-case basis, we recommend our new hybridization scheme for
  future experiments for statistical self-consistency reasons. Second,
  we use the same expression to improve the convergence rate of the
  Blackwell-Rao likelihood estimator, reducing the required number of
  Monte Carlo samples by several orders of magnitude, and thereby
  extend it to high-$\ell$ applications.
\end{abstract}
\keywords{cosmic microwave background --- cosmology: observations --- methods: statistical}

\section{Introduction}
\label{sec:introduction}

The cosmic microwave background (CMB) radiation is one of the most
pristine sources of information about the early Universe available to
us. Since its discovery in 1964 \citep{penzias:1965}, the amount of
information available to us about the CMB has increased at a rapid
pace through series of ground-based, sub-orbital and satellite
experiments. The recently released \emph{Planck} temperature sky maps
\citep{planck:2013_P01} is just the latest example of how the present
challenge in the field of cosmology is one of overabundance rather
than shortage of data. 

To extract cosmological parameters from these ever growing data sets
requires increasingly sophisticated and efficient algorithms, both due
to larger data volumes and to more stringent requirements to
statistical precision. For example, the \emph{COBE}-DMR sky maps
published twenty years ago \citep{smoot:1992} comprised
$\mathcal{O}(10^4)$ pixels, and could be analyzed using exact
brute-force likelihood techniques \citep[e.g.,][]{gorski:1994}, with a
computational scaling of $\mathcal{O}(N_{\textrm{pix}}^3)$. The
\emph{WMAP} sky maps published ten years ago comprised
$\mathcal{O}(10^7)$ pixels \citep{bennett:2003a}, at which point
faster and approximate methods had to be used for parameter estimation
\citep{hivon:2002, verde:2003}. However, for \emph{WMAP} the error
budget was still dominated by cosmic variance on large angular scales
and instrumental noise on small angular scales, and confusion with
Galactic and extra-Galactic emission was minimal, allowing for very
simple component separation methods
\citep{bennett:2003b,hinshaw:2003}. For \emph{Planck}, the total
number of data points in nine frequency bands is
$\mathcal{O}(3\cdot10^8)$, and instrumental noise never dominates the
uncertainties at any angular scales, as small-scale astrophysical
confusion becomes important at multipoles $\ell\gtrsim1500$
\citep{planck:2013_P06}. As a result, an unprecendented study of all
important sources of uncertainty, including instrumental, systematic
and astrophysical, was required for \emph{Planck} to reach its
ambitious goals \citep{planck:2013_P08}.

With the advent of these massive mega-pixel data sets, a number
different analysis strategies have been developed to robustly extract
cosmological parameters with acceptable computational cost. As of
today, the preferred option for full-sky high-resolution experiments
such as \emph{Planck} and \emph{WMAP} is to divide the analysis into
two separate components according to large and small angular scales,
and merge the two at the likelihood level. On large angular scales,
they use a Gibbs sampling based
\citep{jewell:2004,wandelt:2004,eriksen:2004,eriksen:2007}
Blackwell-Rao estimator \citep{chu:2005} that takes into account the
full non-Gaussian structure of the true CMB likelihood, while on small
angular scales, they use faster approaches
\citep[e.g.,][]{hivon:2002,rocha:2011,planck:2013_P08}
coupled to an analytic multivariate Gaussian
(and/or log-normal) likelihood approximation. The computational cost
of this hybrid approach is dominated by spherical harmonics
transforms, and therefore scales as
$\mathcal{O}(N_{\textrm{pix}}^{3/2})$, which is acceptable even for
large data sets. However, there is an unsolved problem associated with
this hybrid approach, and that is how to merge the two likelihood
components into a single all-scale expression; correlations between
the smallest scales in the large-scale likelihood and the largest
scales in the small-scale likelihood should in principle be accounted
for. As of today, no fully satisfactory solution to this exists in the
CMB literature, though various approaches were explored during the Planck
analysis.

Having a computational scaling of
$\mathcal{O}(N_{\textrm{pix}}^{3/2})$, the Gibbs sampling approach
could in principle be employed for all angular scales, thus
eliminating the need for any hybrid approximation. Unfortunately, in
practice this method is in its current implementation limited to low
angular scales for two reasons: First, joint CMB analysis and
component separation is currently implemented in terms of pixel-based
fits of physical foreground models, requiring all frequency bands to
have the same angular resolution, dictated by the coarsest resolution
in a given data set. Second, although the computational scaling for
the Gibbs sampler is acceptable, the prefactor is high. The 2013
\emph{Planck} likelihood employed 100\,000 Gibbs samples in order to
achieve robust Blackwell-Rao convergence, and each of those samples
required $\sim$2000 Conjugate Gradient iterations (and twice as many
spherical harmonic transforms) to converge, for a total cost of
500\,000 CPU hours. Naively scaling this to full \emph{Planck}
resolution suggest a final cost of $\mathcal{O}(10^8)$ CPU hours.

The main result of the present paper is a statistically well motivated
block factorization of the CMB power spectrum likelihood that is
applicable to several of these problems. Specifically, we show that
for sets of random variables that can be arranged sequentially in such
a way that all correlations have a finite range within the sequence,
the full joint probability distribution may be written in terms of
lower-dimensional marginals. The arch-typical example of such a
distribution is a multivariate Gaussian with a strictly banded
covariance matrix, and we therefore call the general (non-Gaussian but
conditionally limited) case also ``banded''. With this statistical
identity ready at hand, we first suggest a statistically
well-motivated likelihood hybridization scheme that takes properly
into account correlations between the low- and high-$\ell$ regimes,
and, second, we show how the convergence rate of the Blackwell-Rao
estimator can be improved by factorizing the full high-dimensional
multivariate posterior into a set of lower-dimensional distributions,
each of which converges much faster than the full distribution. This
approach differs from the direct Gaussianization technique proposed by
\citet{rudjord:2009} in that the underlying probabilistic structure
(e.g., shapes of marginal and $N$-point correlations) is conserved; in
principle, the only modification to the full likelihood enforced by
our new approach is that assumed negligible correlations are
explicitly set to zero.

\section{Factorizing the CMB likelihood}
\label{sec:factorization}

\subsection{Factorization of banded probability distributions}

We begin with a general joint probability density $P(\{\theta\}) =
P(\theta_1, \theta_2, \theta_3, \dots, \theta_n)$ for a set of random
variables, $\theta_k$, with $k=1, 2, 3, \dots, n$.  We choose one
specific sequential ordering of these variables (out of all the
possible orderings), and use the definition of a conditional to write
the joint distribution as a product of univariate conditionals,
\begin{align}
  P(\{\theta\}) = & \quad P(\theta_1, \theta_2, \theta_3, \dots, \theta_n) \nonumber \\
   = & \quad P(\theta_1 | \theta_2, \theta_3, \dots, \theta_n) \nonumber \\
     & \cdot P(\theta_2 | \theta_3, \dots \theta_n) \cdots \nonumber \\ 
     & \cdot P(\theta_{n-1} | \theta_n) \cdot P(\theta_n) \nonumber 
\end{align}
We then assume that our variables only have a conditional probability
dependence on their immediate neighbors in the sequence, i.e., that
the probability distribution is \emph{tri-diagonally} banded,
\begin{align}
  P(\{\theta\}) \approx & \,P(\theta_1 | \theta_2)\cdot\, P(\theta_2 | \theta_3) \cdots \,P(\theta_{n-1} | \theta_n) \cdot \,P(\theta_n) \nonumber \\
  = & \,\frac{P(\theta_1, \theta_2)}{P(\theta_2)}\cdot \frac{P(\theta_2, \theta_3)}{P(\theta_3)}\cdots \frac{P(\theta_{n-1}, \theta_n)}{P(\theta_n)} \cdot P(\theta_n) \nonumber \\
  = & \,\frac{\prod_{k=1}^{n-1}P(\theta_k, \theta_{k+1})}{\prod_{k=2}^{n-1}P(\theta_k)}.
  \label{eq:master}
\end{align}
Thus, this simple derivation shows that a strictly (tri-diagonally)
banded probability distribution may be factorized recursively into a
product of uni- and bivariate marginals.

Before applying this expression to CMB likelihood approximation, we
note that even if the joint probability distribution do not have
correlations exclusively between neighboring variables, it may still
be possible to factorize it, provided at least some correlations may
be ignored. For instance, suppose we can ignore all but the nearest
\emph{two} neighbors; in that case, the joint distribution will
factorize into a product of uni-, bi- and trivariate marginals.

\subsection{Block factorization of the CMB likelihood}
\label{sec:model}

\begin{figure}[t]
  \includegraphics[width=\columnwidth]{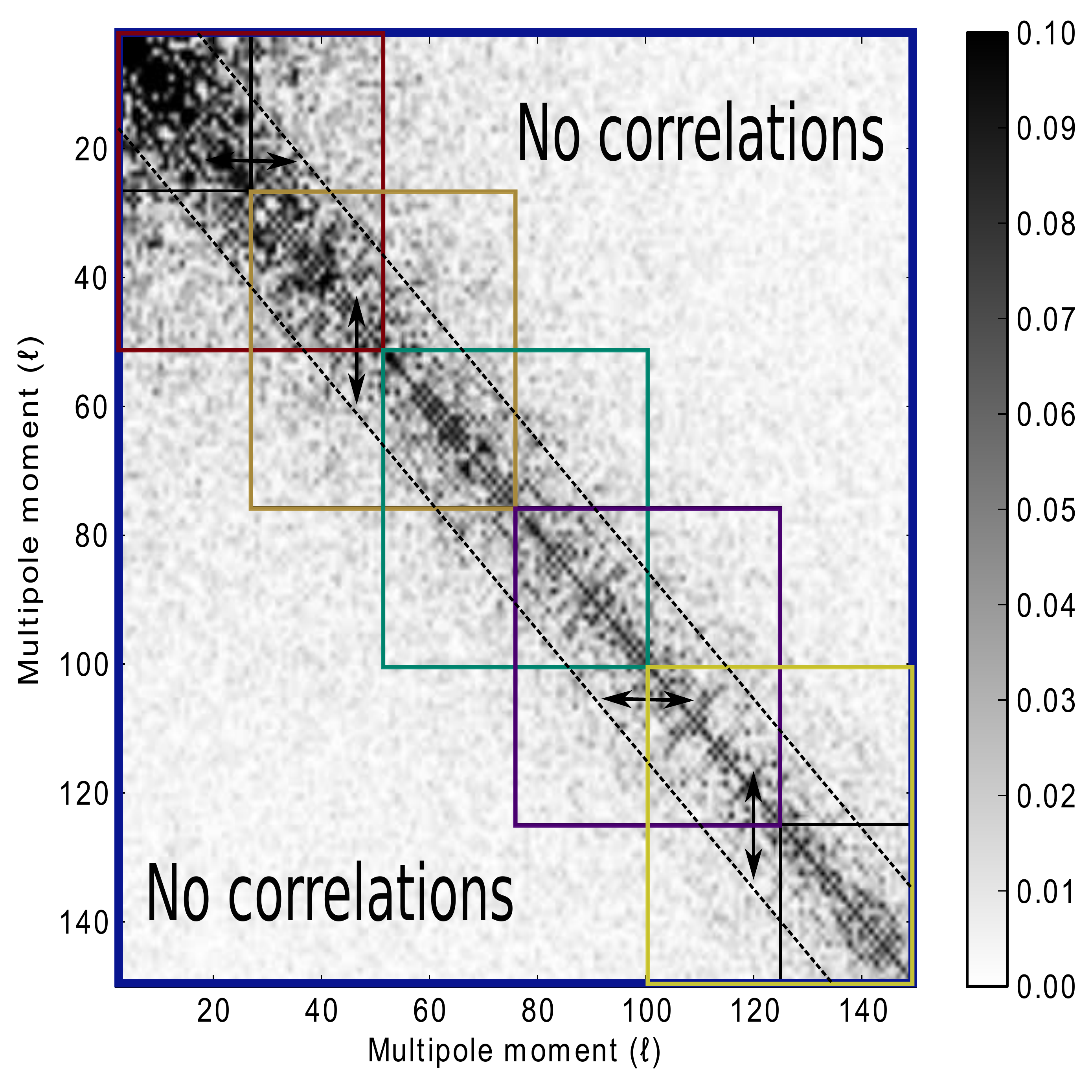}
  \caption{Angular power spectrum correlation matrix, $M_{\ell\ell'}$,
  for the official \emph{Planck} low-$\ell$ CMB data set, estimated by Monte Carlo
    sampling. Note that any two-point correlations are contained
    within a band of $\Delta\ell\sim15$, suggesting that the CMB
    likelihood may be approximated as a banded probability
    distribution. To factorize the CMB likelihood into
    lower-dimensional elements, we partition the full multipole range
    into a set of disjoint blocks such that all non-zero covariance
    elements are embedded within a tri-diagonal block structure,
    indicated here by colored squares.}
  \label{fig:bandmat}
\end{figure}

In its most basic representation, a CMB data set, $\d$, may be
modelled as
\begin{equation}
\d = \s + \n,
\end{equation}
where $\s$ is the true sky signal and $\n$ represents instrumental
noise. Both the signal and noise are usually assumed to be zero-mean
Gaussian variables with covariances $\S$ and $\N$, respectively. 

The noise covariance matrix is typically given by external knowledge
about the instrumental noise characteristics and the scanning strategy
of a given experiment. The signal covariance matrix, on the other
hand, is generally unknown, and must be estimated from the
data. However, given the fact that we only have one observable sky
available, it is impossible to estimate the $N_{\textrm{pix}}^2$
elements in $\S$ from the $N_{\textrm{pix}}$ elements in $\d$ without
imposing strong priors on its structure. The most commonly accepted
prior is simply that the CMB sky is isotropic and homogeneous
\citep[e.g.,][]{planck:2013_P09}. It is therefore convenient to expand
$\s$ in spherical harmonics, such that
\begin{equation}
\s(\hat{\n}) = \sum_{\ell,m} a_{\ell m} Y_{\ell m}(\hat{\n}),
\end{equation}
where $\hat{\n}$ is a unit vector pointing to a given position on the
sky, $Y_{\ell m}$ are the spherical harmonics, and $a_{\ell m}$ are
the corresponding spherical harmonics coefficients. Then the signal
covariance matrix may be written as
\begin{equation}
S_{\ell m, \ell' m'} = \left<a_{\ell m} a^*_{\ell' m'}\right> \equiv
C_{\ell} \delta_{\ell\ell'} \delta_{mm'},
\label{eq:covmat}
\end{equation}
where $C_{\ell}$ is known as the angular power spectrum. 

The main goal of most CMB experiments is precisely to measure the CMB
power spectrum, and the most straightforward way to do so is by
maximum-likelihood estimation. Since we have assumed that both signal
and noise are Gaussian distributed, the CMB power spectrum likelihood
simply reads
\begin{equation}
\mathcal{L}(C_{\ell}) \equiv P(\d|C_{\ell}) \propto
\frac{e^{-\frac{1}{2} \d^t (\S(C_{\ell}) +
    \N)^{-1}\d}}{\sqrt{|\S(C_{\ell}) + \N|}},
\label{eq:likelihood}
\end{equation}
where $\S=\S(C_{\ell})$ is the covariance matrix given in
Equation~\ref{eq:covmat} expressed in pixel domain. Note that
$C_{\ell}$ denotes the set of all power spectrum coefficients, and the
likelihood therefore spans an $\ell_{\textrm{max}}$-dimensional space.

As already mentioned, brute-force evaluation of
Equation~\ref{eq:likelihood} scales computationally as
$\mathcal{O}(N_{\textrm{pix}}^3)$, and is therefore feasible only for
very low angular resolutions. Much of the CMB analysis literature
therefore revolves around finding computationally tractable
approximations to this expression.

In order to build up some intuition about the correlation structure of
$\mathcal{L}(C_{\ell})$, it is useful to plot the correlation matrix
\begin{equation}
M_{\ell\ell'} \equiv \frac{\left< (C_{\ell}-\left<
  C_{\ell}\right>)(C_{\ell'}-\left<
  C_{\ell'}\right>\right>}{\sqrt{\left< (C_{\ell}-\left<
    C_{\ell}\right>)^2\right>\left<(C_{\ell'}-\left< C_{\ell'}\right>)^2\right>}}.
\end{equation}
Figure~\ref{fig:bandmat} shows this matrix for the official \emph{Planck} low-$\ell$ CMB data, as evaluated from $200,000$ Monte Carlo samples generated 
with a CMB Gibbs sampler
\citep{eriksen:2007}. In this case, there are significant correlations
between all elements at $\ell\lesssim20$, while at $\ell\gtrsim50$ any
correlations are well contained inside a band of $\Delta\ell=15$; any
correlations beyond $\Delta\ell\gtrsim30$ are well below 1\,\%. Higher-order
correlations are significantly smaller than these two-point
correlations. 

\begin{figure*}[t]
  \begin{center}
    \includegraphics[width=0.33\textwidth]{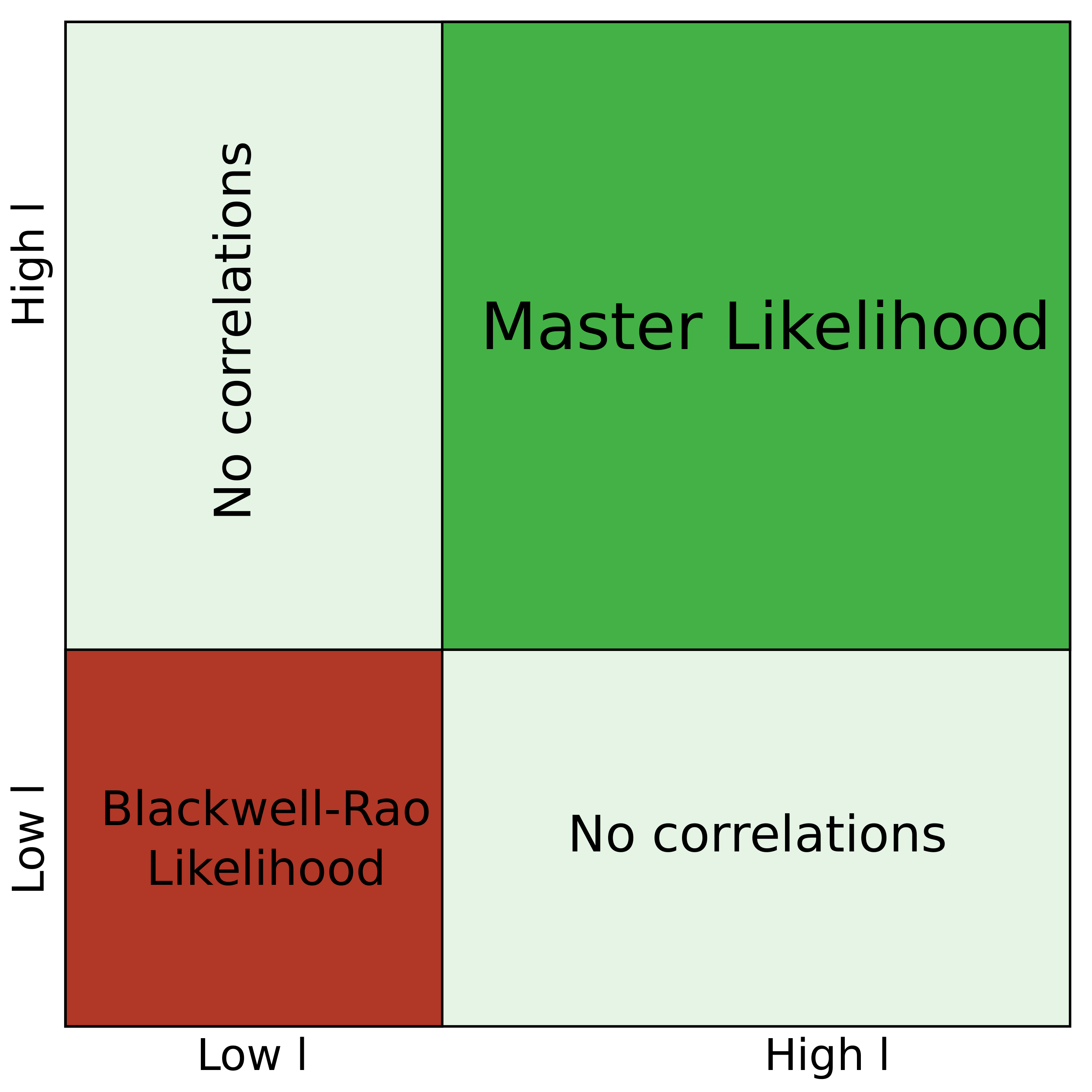}
    \includegraphics[width=0.33\textwidth]{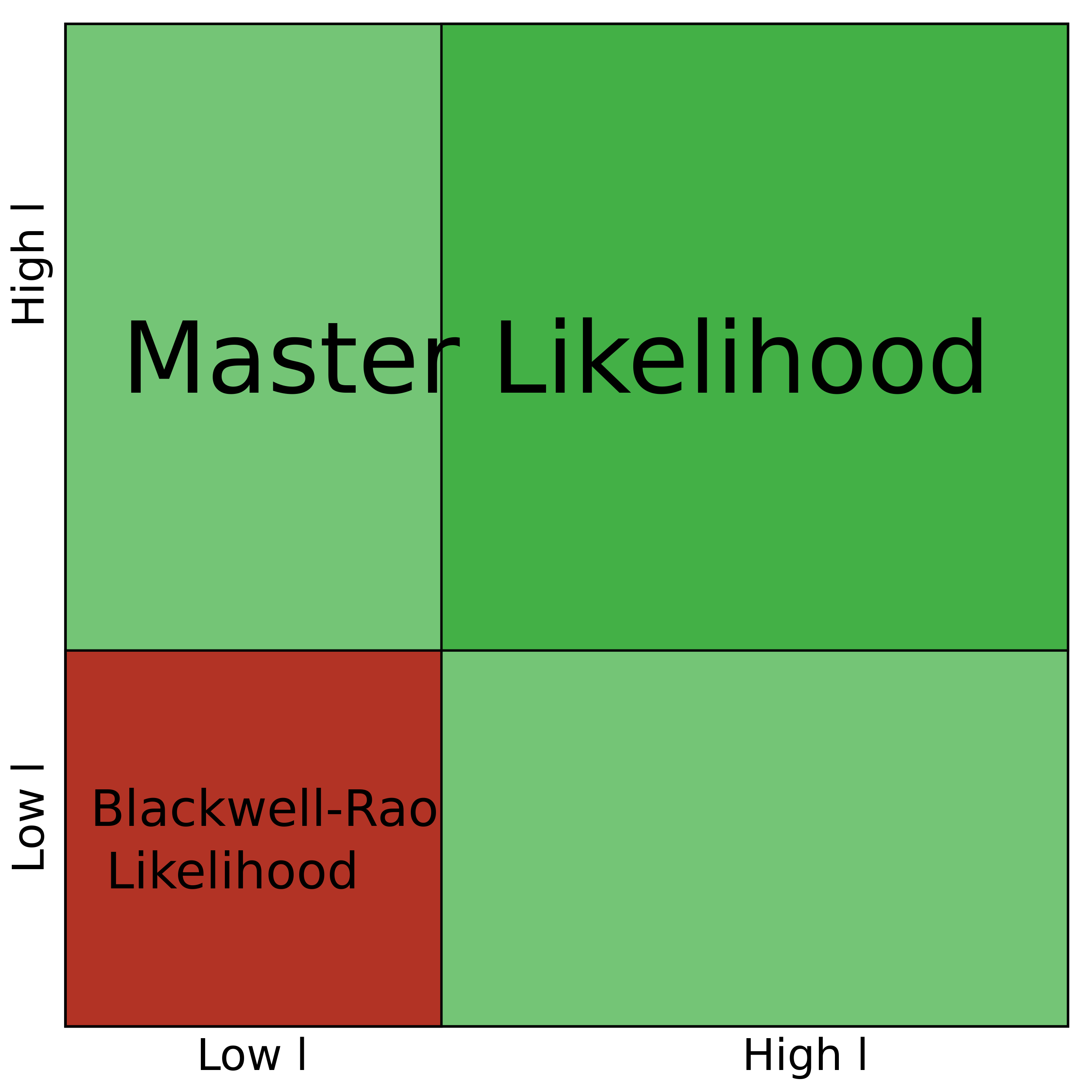}
    \includegraphics[width=0.33\textwidth]{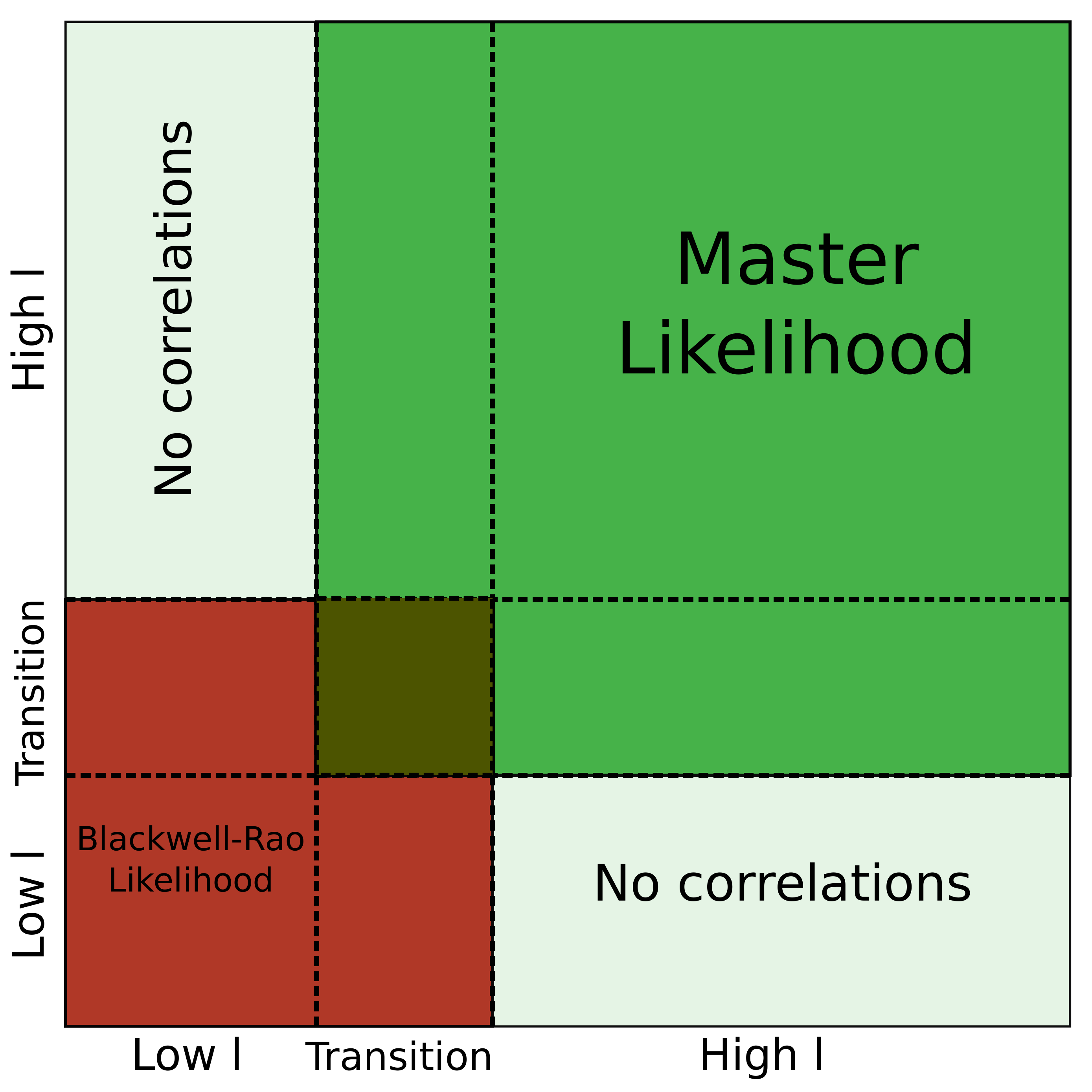}
  \end{center}
  \caption{Schematic overview of the three hybridization schemes
  discussed in the text. The left panel illustrates a sharp transition
  between the low- (Blackwell-Rao) and high-$\ell$ (MASTER)
  likelihood, as currently adopted by \emph{Planck}. The middle panel
  illustrates the \emph{WMAP} approach, which includes the
  off-diagonal elements between the low- and high-$\ell$ regions in
  the high-$\ell$ likelihood estimator. The right panel illustrates
  the new estimator proposed in this paper, in which correlations are
  accounted for through an transition region that is sufficiently wide
  to include all non-negligible correlations between the low- and
  high-$\ell$ regions. To avoid double-counting of the diagonal
  elements, the total log-likelihood is corrected by the
  log-likelihood including elements within the transition region only. }
  \label{fig:regions}
\end{figure*}

For typical sky cuts and instrumental noise characteristics, the basic
CMB likelihood can therefore be approximated as a banded probability
distribution with a bandwidth of $\Delta\ell\lesssim15$, and can
therefore in principle be factorized by
Equation~\ref{eq:master}. However, as currently written this
expression only applies to a strictly tri-diagonal covariance
matrix. To circumvent this problem, we therefore introduce an
auxiliary block structure that embeds all non-negligible elements
within a larger tri-diagonal structure, as illustrated by the colored
blocks in Figure~\ref{fig:bandmat}. That is, we define a set of
multipole blocks such that $\theta_1 =
\{C_{\ell_\textrm{min}},\ldots,C_{\ell_1}\}$, $\theta_2 =
\{C_{\ell_1+1},\ldots,C_{\ell_2}\}$, \ldots, $\theta_n =
\{C_{\ell_{n-1}+1},\ldots,C_{\ell_\textrm{max}}\}$. Thus, each
univariate marginal in Equation~\ref{eq:master} is replaced with a
multivariate distribution of dimension $\ell_{i}-\ell_{i-1}$, and each
bivariate marginal is replaced with a multivariate distribution of
dimension $\ell_{i}-\ell_{i-2}$. This block-wise factorization
constitutes the main result of this paper, and in the following
sections we will apply this to two concrete problems in CMB likelihood
estimation.

\begin{deluxetable*}{lcccccc}[h!t]
\tablewidth{0pt}
\tablecaption{\label{tab:trans_tab}Summary of cosmological parameters
  derived with three different hybridization schemes; the original
  \emph{WMAP} approach including off-diagonal elements in the inverse
  covariance matrix (\emph{second column}); a sharp transition at
  $\ell_{\textrm{trans}}=32$ (\emph{third column}); and the new
  approach implementing a transition region between $\ell=21$ and 32
  (\emph{fifth column}). The fourth and sixth columns show the
  relative shifts
  with respect to the \emph{WMAP} approach measured in units of $\sigma$. }
\tablecomments{The confidence intervals are 1 $\sigma$, and the best-fit points are the marginalised means of the parameters.}
\tablecolumns{6}
\tablehead{ & \textbf{Default WMAP}& \multicolumn{2}{c}{\textbf{Sharp transition}} &
  \multicolumn{2}{c}{\textbf{Transition region}} \\
& Constraint & Constraint  & Deviation ($\sigma$) & Constraint  & Deviation ($\sigma$)}
\startdata
$\Omega_b h^2$              & $0.0225\pm 0.0006$ & $0.0225\pm 0.0006$ & $0.02$ & $0.0225\pm 0.0006$ & $0.02$ \\
$\Omega_m h^2$              & $0.111 \pm 0.005 $ & $0.111 \pm 0.005 $ & $0.01$ & $0.112 \pm 0.006 $ & $0.05$ \\
$\theta$                    & $1.039 \pm 0.003 $ & $1.039 \pm 0.003 $ & $0.04$ & $1.039 \pm 0.003 $ & $0.05$ \\
$\tau$                      & $0.088 \pm 0.015 $ & $0.088 \pm 0.015 $ & $0.04$ & $0.088 \pm 0.015 $ & $0.05$ \\
$n_{s}$                     & $0.969 \pm 0.013 $ & $0.969 \pm 0.014 $ & $0.03$ & $0.968 \pm 0.014 $ & $0.06$ \\
$\textrm{log}[10^{10} A_s]$ & $3.08  \pm 0.04  $ & $3.08  \pm 0.03  $ & $0.03$ & $3.08  \pm 0.04  $ & $0.05$
\enddata
\end{deluxetable*}

\section{Accurate hybrid CMB likelihood estimation}
\label{sec:hybrid}

As already mentioned, both \emph{Planck} and \emph{WMAP} have adopted
so-called "hybrid" likelihood approximations, combining a Gibbs
sampling based Blackwell-Rao estimator at large angular scales with a
Gaussian (and/or log-normal) pseduo cross-spectrum approximation at
small angular scales. These two components are merged into a single
expression at the log-likelihood level. The \emph{Planck} likelihood
simply adds the two log-likelihoods \citep{planck:2013_P08}, adopting
a so-called ``sharp transition'' between the low- and high-$\ell$
regimes, schematically illustrated in the left panel of
Figure~\ref{fig:regions}. This is the simplest possible approach, and
assumes that any correlations across the transition multipole are
negligible. The \emph{WMAP} likelihood makes a different choice, by
including the off-diagonal terms between the low- and high-$\ell$
blocks in the (Gaussian plus log-normal) high-$\ell$ likelihood, as
illustrated in the middle panel of Figure~\ref{fig:regions}.

In this section, we introduce a new and statistically better motivated
approach than either of two employed by \emph{Planck} and \emph{WMAP},
taking advantage of the block factorization derived in
Equation~\ref{eq:master}. The first step in our approach is to
partition the full multipole range between $\ell_{\textrm{min}}$ and
$\ell_{\textrm{max}}$ into three disjoint regions,
$L=\{\ell_{\textrm{min}}, \ldots, \ell_{\textrm{low}}\}$,
$T=\{\ell_{\textrm{low}}+1, \ldots, \ell_{\textrm{high}}-1\}$ and
$H=\{\ell_{\textrm{high}}, \ldots, \ell_{\textrm{max}}\}$,
corresponding to a low-$\ell$ region, a transition region and a
high-$\ell$ region, respectively. The width of the transition region
is chosen to be at least as wide as the effective bandwidth of the
$C_{\ell}$ covariance matrix (see Figure~\ref{fig:bandmat}). With this
partitioning, we now specialize Equation~\ref{eq:master} to the case
with $n=3$ regions;
\begin{equation}
\log\mathcal{L}(C_{\ell}) = \log\mathcal{L}(L,T) +
\log\mathcal{L}(T,H) - \log\mathcal{L}(T).
\label{eq:hybrid}
\end{equation}
Note that this approximation is exact under the assumption of
vanishing correlations between the low- and high-$\ell$ regions, which
can be ensured simply by letting the transition region be sufficiently
wide. This estimator is schematically illustrated in the right panel
of Figure~\ref{fig:regions}. 

Equation~\ref{eq:hybrid} has a simple intuitive interpretation: The
log-likelihood is simply the sum of a low- and a high-$\ell$
contribution, defined such that they overlap over a sufficiently wide
multipole range that all non-negligible correlations are
included. However, because the diagonal block in the transition region
is included twice, both by the low- and the high-$\ell$ likelihood,
one must subtract the corresponding marginal for the transition region
once to avoid double-counting (this is also an immediate consequence of
equation 1, under the assumption that $p(L|T, H) = p(L|T) = p(L, T) / P(T)$, 
i.e. the low-$\ell$ region is conditionally independent of the high-$\ell$
region given the transition region). Note that any estimator for the
transition likelihood may be used for the correction term, typically
by extracting the relevant range from either the low- or the
high-$\ell$ likelihoods.

\begin{figure*}[t]
  \begin{center}
    \includegraphics[width=\linewidth]{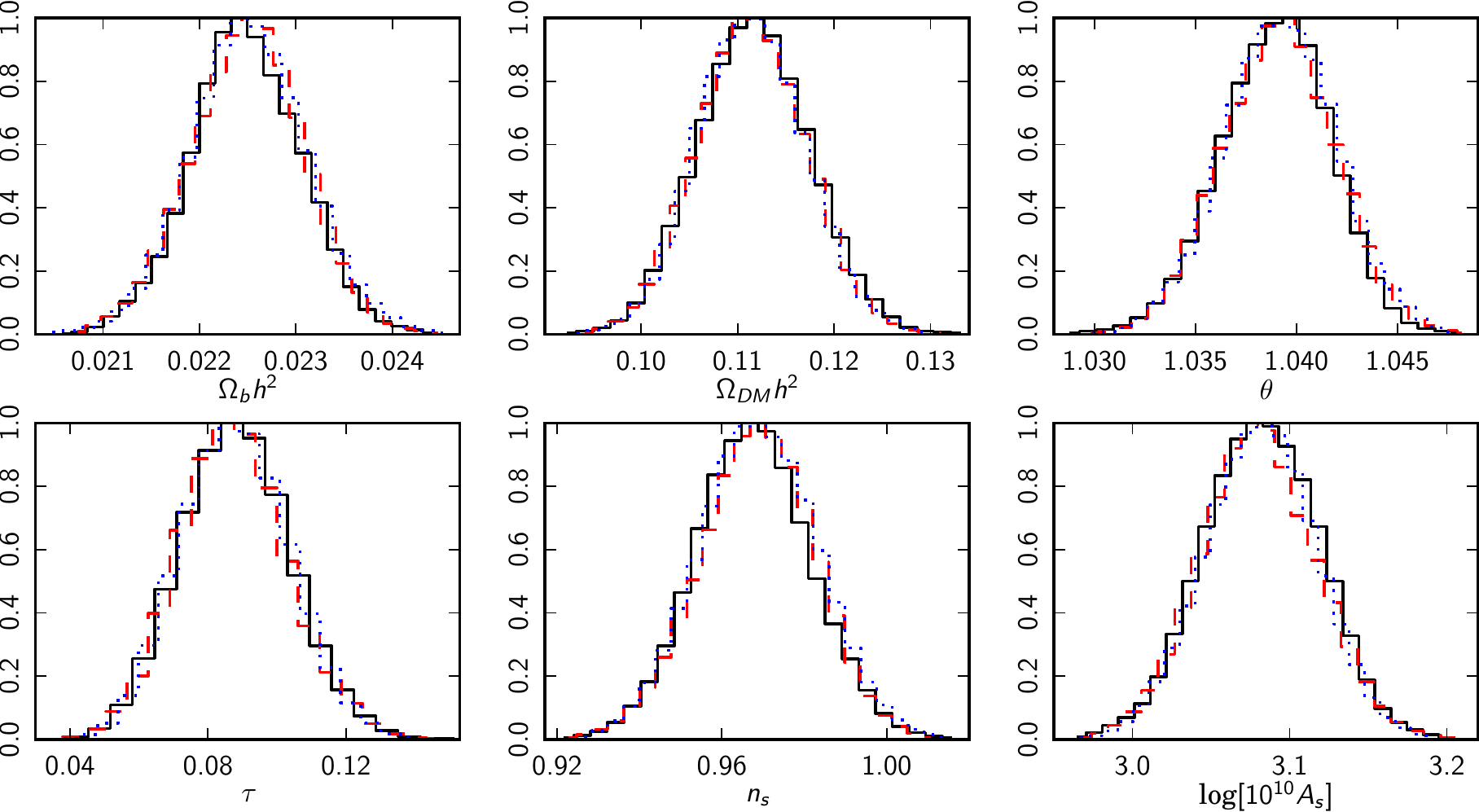}
  \end{center}
  \caption{Comparison of best-fit parameters derived by CosmoMC from
    \emph{WMAP} using likelihood approximations based on the new
    hybrid estimator presented in this paper (\emph{solid black line}); the
    \emph{WMAP} approach including off-diagonal elements in the
    inverse covariance matrix (\emph{dashed red line}); and a sharp
    transition between the low- and high-$\ell$ regions (\emph{dotted blue
    line}).}
  \label{fig:trans_plot}
\end{figure*}

To assess the importance of the specific strategy adopted for
hybridization, we modify the (7-year) \emph{WMAP} likelihood to
include each of the three solutions, and derive constraints on the
standard $\Lambda$CDM model using \emph{WMAP} data only. The
transition multipole is set to $\ell_{\textrm{trans}}=32$ for the
sharp transition case, whereas the transition region is defined as
$\ell=\{21,\ldots,32\}$ for the new hybrid scheme. The \emph{WMAP}
Blackwell-Rao estimator is used both for the low-$\ell$ and the
transition regions in the latter case. We adopt $\Omega_b h^2,
\Omega_m h^2, \theta, \tau, n_s,$ and $\log(10^{10}A_s)$ as our
primary parameters, and adopt CosmoMC \citep{lewis:2002} as our MCMC
engine. The resulting one-dimensional marginals are shown in
Figure~\ref{fig:trans_plot} for all three cases, and posterior mean
summary statistics are given in Table \ref{tab:trans_tab}.

With a largest relative difference between any two cases of
$0.06\sigma$, these results demonstrate that the standard
six-parameter $\Lambda$CDM model is highly robust with respect to
assumptions about the correlations across the transition
regime. Similar conclusions were found when performing an identical
analysis for the the recently released \emph{Planck} likelihood
\citep{planck:2013_P08}, and this motivated the choice of a sharp
transition for that particular implementation. For future experiments
and analyses we nevertheless recommend the hybrid approach presented
here, for two main reasons. First, our expression provides a
statistically well motivated solution whose validity may be monitored
directly through the $C_{\ell}$ covariance matrix; without the same
level of statistical rigour, detailed simulations are more critical
for the other two approaches, and these should in principle be
repeated both when the data set or the parametric model is
changed. Second, this expression is implementationally trivial once
both low- and high-$\ell$ likelihoods are available, and there is
therefore no practical reason for not including these correlations,
even if their impact may be small.

\section{Faster Blackwell-Rao convergence}
\label{sec:br}

\subsection{Review of the Blackwell-Rao estimator}

As mentioned in Section~\ref{sec:introduction}, both the \emph{Planck}
and \emph{WMAP} low-$\ell$ likelihoods \citep{planck:2013_P08,
  hinshaw:2012} employs a specific Blackwell-Rao (BR) estimator to
produce an accurate likelihood approximation that accounts for all
correlations and non-Gaussian structures \citep{chu:2005}. The main
advantages of this estimator are 1) computational speed, 2) implementational
simplicity, and 3) support for seamless marginalization over
systematic effects and component separation errors through Gibbs
sampling \citep{eriksen:2007}. 

This estimator may be explained intuitively as follows: Suppose it is
possible to construct an experiment that provides a perfect full-sky
noiseless image of the CMB sky, $\d=\s$. For that experiment, the only
source of uncertainty on $C_{\ell}$ is cosmic variance, and the exact
CMB likelihood in Equation~\ref{eq:likelihood} reduces to an inverse
Gamma distribution,
\begin{equation}
\mathcal{L}_0(C_{\ell}) \propto
\frac{e^{-\frac{1}{2} \s^t
    \S(C_{\ell})^{-1}\s}}{\sqrt{|\S(C_{\ell})|}} \propto \prod_{\ell}
\sigma_{\ell}^{-\frac{2\ell-1}{2}}\frac{e^{\frac{2\ell+1}{2}\frac{\sigma_{\ell}}{C_{\ell}}}}{C_{\ell}^{\frac{2\ell+1}{2}}}.
\label{eq:BR0}
\end{equation}
Here we have defined $\sigma_{\ell} \equiv \frac{1}{2\ell+1}
\sum_{\ell=-m}^m |a_{\ell m}|^2$ to be the realization specific power
spectrum of $\mathbf{s}$. 

However, for any real experiment there are additional sources of
uncertainty beyond cosmic variance, for instance from instrumental
noise and foreground contamination, and $P(\s|\d)$ is no longer a
delta function. In order to account for this additional uncertainty,
one must weight the ideal likelihood in Equation~\ref{eq:BR0} with
respect to $P(\s|\d)$,
\begin{equation}
\mathcal{L}_{\textrm{BR}}(C_{\ell}) = \int d\s \,\mathcal{L}_0(C_{\ell})\,
P(\s|\d).
\end{equation}
At first glance, this integral appears difficult to evaluate, as it
involves millions of degrees of freedom. However, this is precisely
where the CMB Gibbs sampler enters the picture. As explained in detail
by \citet{jewell:2004, wandelt:2004, eriksen:2004, eriksen:2007}, the
output from this algorithm is a set of samples drawn directly from
$P(\s|\d)$, accounting for both instrumental noise and foreground
errors. Thus, the integral can be simply evaluated by Monte Carlo
integration as a sum over these samples,
\begin{equation}
\mathcal{L}_{\textrm{BR}}(C_{\ell}) \approx \sum_{i=1}^{N_{\textrm{samp}}} \prod_{\ell=\ell_{\textrm{min}}}^{\ell_{\textrm{max}}}
{\sigma^i_{\ell}}^{\frac{2\ell-1}{2}}\frac{e^{\frac{2\ell+1}{2}\frac{\sigma^i_{\ell}}{C_{\ell}}}}{C_{\ell}^{\frac{2\ell+1}{2}}}.
\label{eq:BR2}
\end{equation}
This is the CMB power spectrum Blackwell-Rao estimator, which is
guaranteed to converge to the true likelihood in the limit of
$N_{\textrm{samp}}\rightarrow\infty$.

\subsection{Lifting the ``curse of dimensionality'' \\by block factorization}

\begin{figure}[t]
  \begin{center}
    \includegraphics[width=\columnwidth]{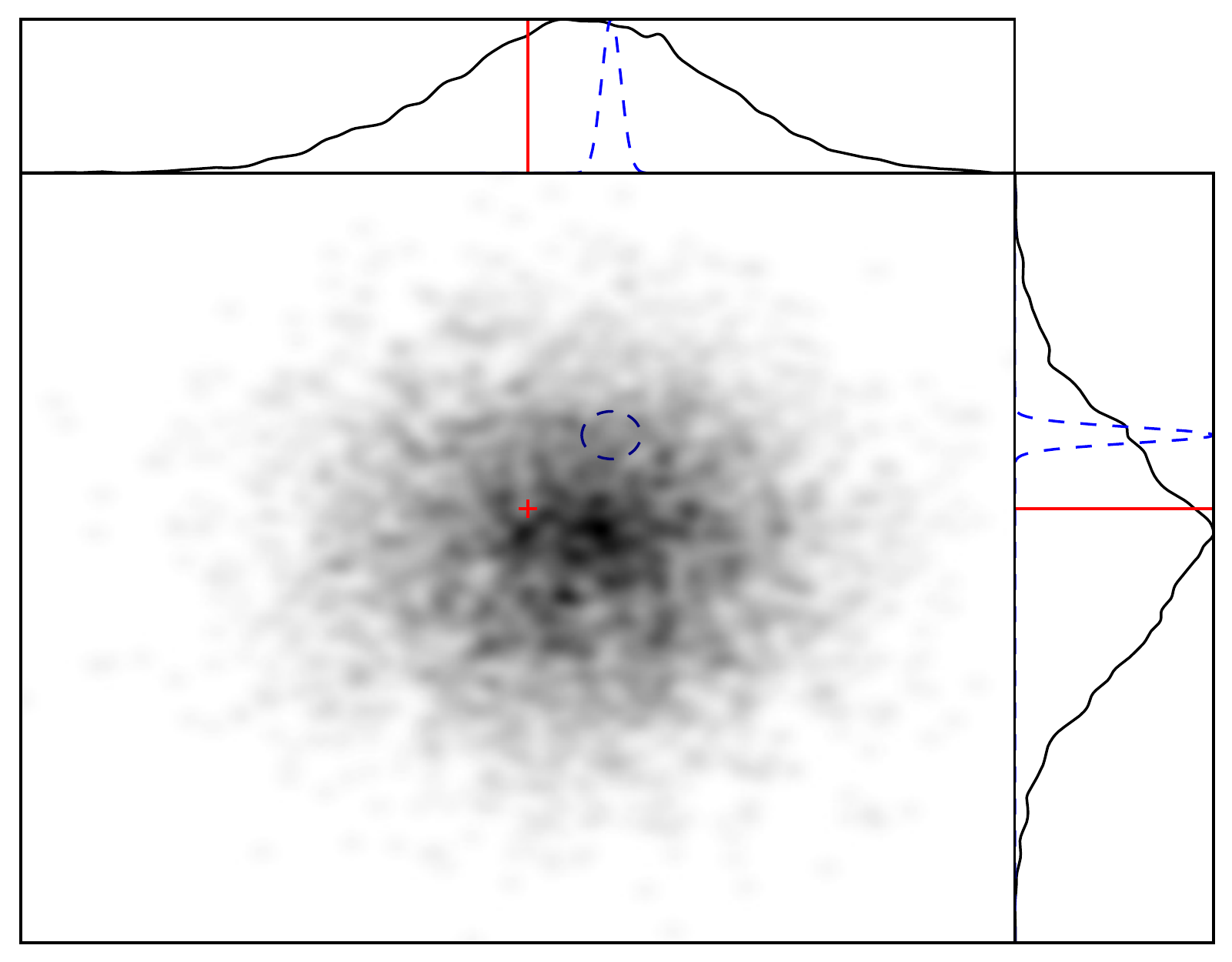}
  \end{center}
  \caption{Illustration of the ``curse of dimensionality''. The
    Blackwell-Rao estimator builds up a smooth histogram from a finite
    set of Monte Carlo samples by assigning a distribution (or kernel)
    to each sample. The number of samples required to reach
    convergence is proportional to the ratio between the volume of the
    kernel (\emph{blue}) and the volume of the full distribution
    (\emph{black}). If this ratio is $r<1$ in one dimension (top and
    left panels), it is $r^2$ in two dimensions (central panel), and
    $r^n$ in $n$ dimensions. This implies that the number of Monte
    Carlo samples required to reach convergence for the CMB BR
    estimator scales exponentially with $\ell_{\textrm{max}}$. The evaluation
    of the 2-d likelihood at a specific point in parameter space
    (\emph{red cross}) will be much more sensitive to the number of
    samples than the corresponding evaluations in the respective marginalized
    parameter spaces (\emph{red lines}).}
  \label{fig:dimensions}
\end{figure}

While the Blackwell-Rao estimator is guaranteed to converge to the
correct answer, it is not obvious how fast it does so, as measured in
terms of number of samples required for convergence,
$N_{\textrm{samp}}$. Further, since the computational cost of a single
Gibbs sample is typically on the order of several CPU hours
\citep{eriksen:2004}, depending on the angular resolution and/or
signal-to-noise ratio of the data set under consideration, it is
important to understand this scaling before attempting a full-scale
analysis. Indeed, \citet{chu:2005} showed that $N_{\textrm{samp}}$
scales exponentially with $\ell_{\textrm{max}}$, effectively limiting
its operational range to $\ell_{\textrm{max}}\approx50$--70. The main
goal of the present section is to improve on this limit, and extend
the BR estimator to high $\ell$'s.

To understand the origin of the exponential scaling, we show in
Figure~\ref{fig:dimensions} a simple two-dimensional Gaussian
distribution mapped by a Monte Carlo sampler. The top and left panels
show the respective one-dimensional marginals. The Blackwell-Rao
estimator establishes a smooth approximation to these distributions by
assigning a kernel of finite width to each individual Monte Carlo
sample (illustrated by blue contours/Gaussians) before taking the
average over all samples. Suppose now that the width of the
one-dimensional kernel is 10\% of the width of the marginal
distribution; in that case, one needs $\sim$10 samples in order to
cover the marginal once. In two dimensions, however, one needs
$\sim$$10^2$ samples to cover the full joint distribution once, since
the ratio now is only 10\% in each of the two directions. More
generally, in $n$ dimensions one would need $\sim$$10^n$ samples. This
is a variation of the well-known ``curse of dimensionality'', which
says that the number of points required to cover an $n$-dimensional
space scales exponentially with $n$.

\begin{figure*}[t]
  \begin{center}
    \includegraphics[width=\linewidth]{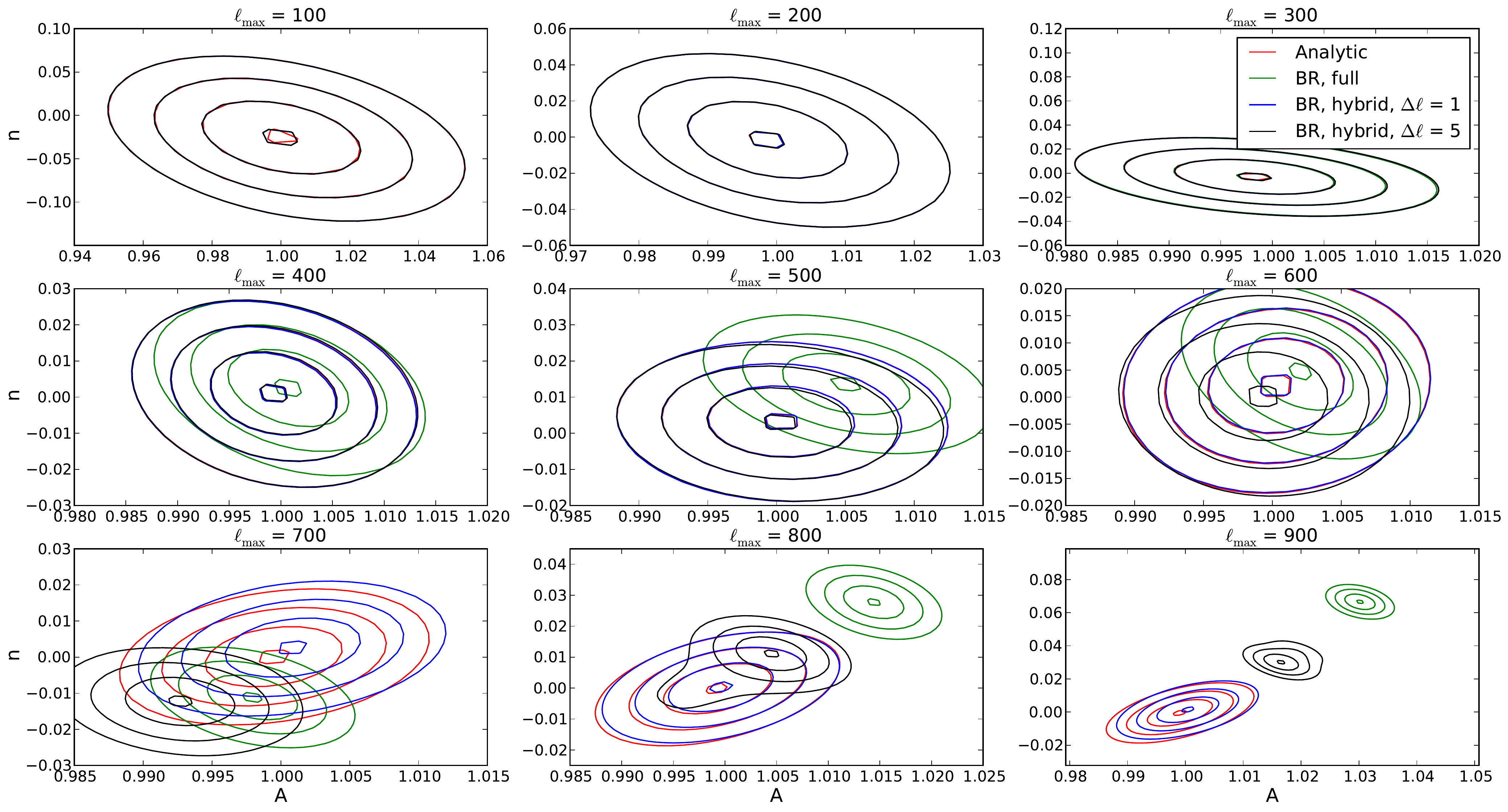}
  \end{center}
  \caption{Comparison of four different methods of evaluating a simple amplitude-tilt likelihood for a full-sky simulation: The analytic case, the full Blackwell-Rao case, and two versions of the hybrid likelihood described in this paper - with $\Delta\ell = 1$ and $5$, respectively.}
  \label{fig:hirespoc}
\end{figure*}

The BR estimator given in Equation~\ref{eq:BR2} converges well up to
$\ell\approx30$ with only a few thousand samples for \emph{WMAP}
\citep{chu:2005}, while for \emph{Planck} it is found to be robust up
to $\ell\approx70$ with 100\,000 samples \citep{planck:2013_P08}. To
extend to even higher $\ell$'s by brute force would soon require a
prohibitively large number of samples, as the computational cost for
the Gibbs sampling step of the latter case is already half a million
CPU hours.

Fortunately, the block factorization presented in
Section~\ref{sec:factorization} may be used to define an alternative
and computationally much cheaper algorithm:
\begin{enumerate}
\item Partition the full $\ell_{\textrm{max}}$-dimensional
  $\mathcal{L}(C_{\ell})$ into a sequence of lower-dimensional blocks,
  $r_k$, for instance of width $\Delta\ell$.
\item Use the standard BR estimator to estimate the marginal
  likelihood for each block and each neighboring set of two blocks. 
\item Merge these block marginals into a single all-$\ell$ estimator
  through the block factorization in Equation~\ref{eq:master}.
\end{enumerate}
Thus, our new likelihood approximation can be written succinctly on
the following form,
\begin{equation}
  \mathcal{L}(C_\ell) \approx \frac{\prod_{k=1}^{n-1}\mathcal{L}_{\textrm{BR}}(r_k, r_{k+1})}{\prod_{k=2}^{n-2}\mathcal{L}_{\textrm{BR}}(r_k)}.
  \label{eq:splitlike}
\end{equation}
Note that all the likelihood evaluations on the right side of this
expression involve a maximum of $2\Delta\ell - 1$ dimensions, as
opposed to $\ell_{\mathrm{max}} - \ell_{\mathrm{min}} + 1$ for the
full joint BR estimator, effectively lifting the curse of
dimensionality. 

\subsection{Accuracy and convergence}
\label{sec:bran}

\subsubsection{Methodology}
Before the block factorized BR estimator can be used for real
analysis, it is necessary to assess its accuracy and convergence
properties. To this aim, we analyze two different simulations with the
above machinery, adopting the convergence analysis methodology of
\citet{chu:2005}, but implementing a few minor changes to improve the
reliability of the convergence statistics. Monte Carlo samples are
produced with Commander \citep{eriksen:2004,eriksen:2007}.

The first simulation consists of a full-sky high-resolution
($N_{\textrm{side}}=512$, $\ell_{\textrm{max}}=1024$, 14' Gaussian
beam) data set with uniform noise (65\,$\mu\textrm{K}$ RMS per
pixel). The main advantage of this case is that the $C_{\ell}$
likelihood (Equation~\ref{eq:likelihood}) factorizes in $\ell$, and
can be evaluated analytically,
\begin{equation}
\mathcal{L}_{\textrm{ideal}}(C_{\ell}) \propto \prod_{\ell}
\frac{e^{-\frac{2\ell+1}{2}\frac{\hat{\sigma}_{\ell}}{(C_{\ell}+N_{\ell})}}}{(C_{\ell}+N_{\ell})^{\frac{2\ell+1}{2}}},
\label{eq:ideal_like}
\end{equation}
where $\hat{\sigma}_{\ell}$ is the angular power spectrum of the noisy
sky map, and $N_{\ell}$ is the ensemble averaged noise power spectrum.
The second simulation consists of a low-resolution
($N_{\textrm{side}}=32$, $\ell_{\textrm{max}}=95$, $4^{\circ}$ FWHM
Gaussian beam) data set with the \emph{WMAP} KQ85 sky cut imposed,
removing 25\,\% of the sky. White noise of 5\,$\mu\textrm{K}$ RMS is
added to each pixel, resulting in a signal-to-noise of unity at
$\ell\approx70$. The main purpose of this simulation is to study the
effect of correlations between different multipoles arising from the sky cut
through comparison
with brute-force pixel-space likelihood evaluation. However, because
of the brute-force evaluations, this case is necessarily limited to
low angular resolution.

The CMB signal is drawn from a Gaussian distribution with a covariance
given by the best-fit \emph{WMAP} $\Lambda$CDM power spectrum,
$C_\ell^{\mathrm{ref}}$ \citep{hinshaw:2012}. In each case, we fit a
two-parameter amplitude-tilt ($A$--$n$) model on the form
\begin{equation}
  C_\ell(A, n) = A \,\left(
  \frac{\ell}{\ell_{0}} \right)^n\, C_\ell^{\mathrm{ref}},
  \label{eq:andefined}
\end{equation}
where $\ell_0=\ell_{\textrm{max}}/2$, simply by
mapping out $\mathcal{L}(A,n)$ over a two-dimensional grid. 
For $\ell_{\mathrm{min}} = 2$, this choice of pivot multipole ensures a low
degree of correlation between $A$ and $n$.

To assess both convergence and accuracy, we adopt the following
measure of difference between two likelihoods, $\mathcal{L}_1$ and
$\mathcal{L}_2$ \citep{chu:2005},
\begin{equation}
  q = \int\left|\mathcal{L}_1(A, n) - \mathcal{L}_2(A, n)\right|\,dA\,dn.
  \label{eq:qdef}
\end{equation}
One can show that if $\mathcal{L}_1$ and $\mathcal{L}_2$ are two
bivariate Gaussian distributions with the same covariance matrix,
$\mathbf{\Sigma}$, but different means, $\vec{\mu}_1$ and
$\vec{\mu}_2$, then
\begin{equation}
  q = \Phi\left( \frac{1}{2\sqrt{2}}\sqrt{(\vec{\mu}_1 - \vec{\mu}_2)\mathbf{\Sigma}^{-1}(\vec{\mu}_1 - \vec{\mu}_2)} \right), 
\end{equation}
where $\Phi$ is the cumulative standard normal
distribution function. From this, one finds that a $0.1\sigma$ shift
in a Gaussian distribution corresponds to $q\sim0.05$. In the
following, we therefore define two distributions to agree if $q<0.05$.

For the accuracy assessment, we simply compare the block factorized BR
likelihood with the exact case. Convergence assessment, however, is
done by drawing two disjoint sample subsets from the full set of
available Monte Carlo samples, compute the BR estimator from each
subset, and compare the resulting likelihoods. We then increase the
number of samples in the two subsets, $N_{\textrm{samp}}$, until $q$
is consistently lower than 0.05 even when adding 100 additional
samples; the latter criterion is imposed in order to avoid chance
agreement. Finally, we repeat this calculation a certain number of times with
different sample subsets (but drawn from the same full sample set),
and report the median of the resulting values of $N_{\textrm{samp}}$
as the final estimate of the number of samples required for
convergence.

\subsubsection{Results}

Figure~\ref{fig:hirespoc} shows $\mathcal{L}(A,n)$ evaluated from the
high-resolution full-sky simulation for nine different values of
$\ell_{\textrm{max}}$ with four different likelihood expressions;
analytic, standard BR, and two variations of the block-factorized BR
estimator. A total of $N_{\textrm{samp}}=28,000$ samples are included in
the two latter, a choice that is set to highlight the fundamental
difference between the various cases. In particular, since there are
no correlations between any multipoles in this case, all four
approaches are in principle exact, and the only difference among the
four cases are their relative convergence rates.

For $\ell_{\textrm{max}} \le 300$, we see that all four estimator
agree to very high accuracy. However, from $\ell_{\textrm{max}} \ge
400$ the full-range BR likelihood starts to diverge. 
At $\ell_{\textrm{max}} = 900$, it is separated from the
analytic result by more than $15\sigma$. In this case, the sum in
Equation~\ref{eq:BR2} is strongly dominated by the one sample that
happens to have the lowest power spectrum scatter about some best-fit
mode, and the resulting distribution is simply an imprint of the
cosmic variance kernel (Equation~\ref{eq:BR0}) for that sample.

\begin{figure}[t]
  \begin{center}
    \includegraphics[width=\linewidth]{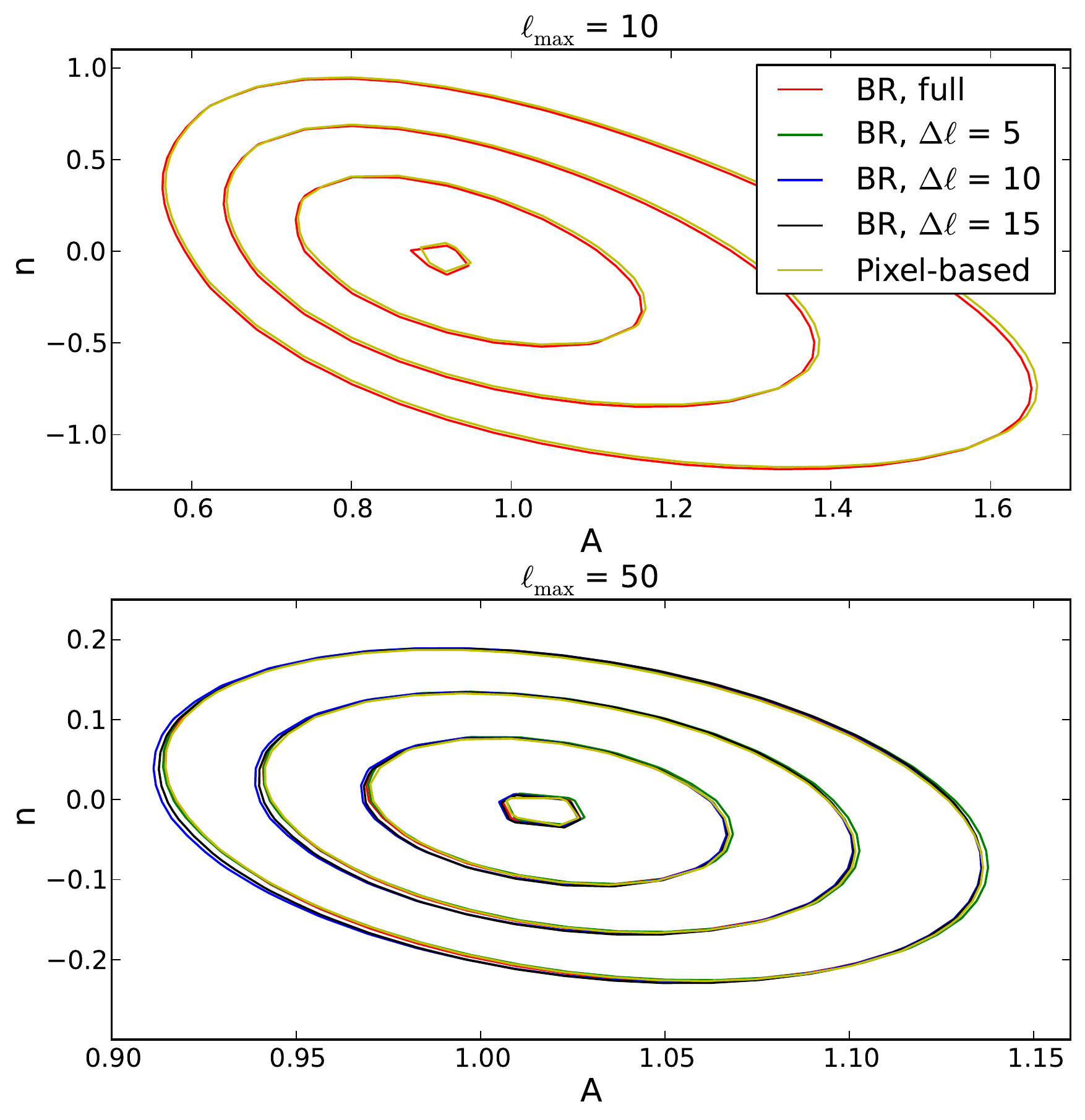}
  \end{center}
  \caption{Comparison of five different methods of evaluating a simple amplitude-tilt likelihood for a cut-sky simulation: The pixel-based case, the full Blackwell-Rao case, and three versions of the hybrid likelihood described in this paper - with $\Delta\ell = 5$, $10$, and $15$, respectively.}
  \label{fig:maskedpoc}
\end{figure}

\begin{figure}[t]
  \begin{center}
    \includegraphics[width=\linewidth]{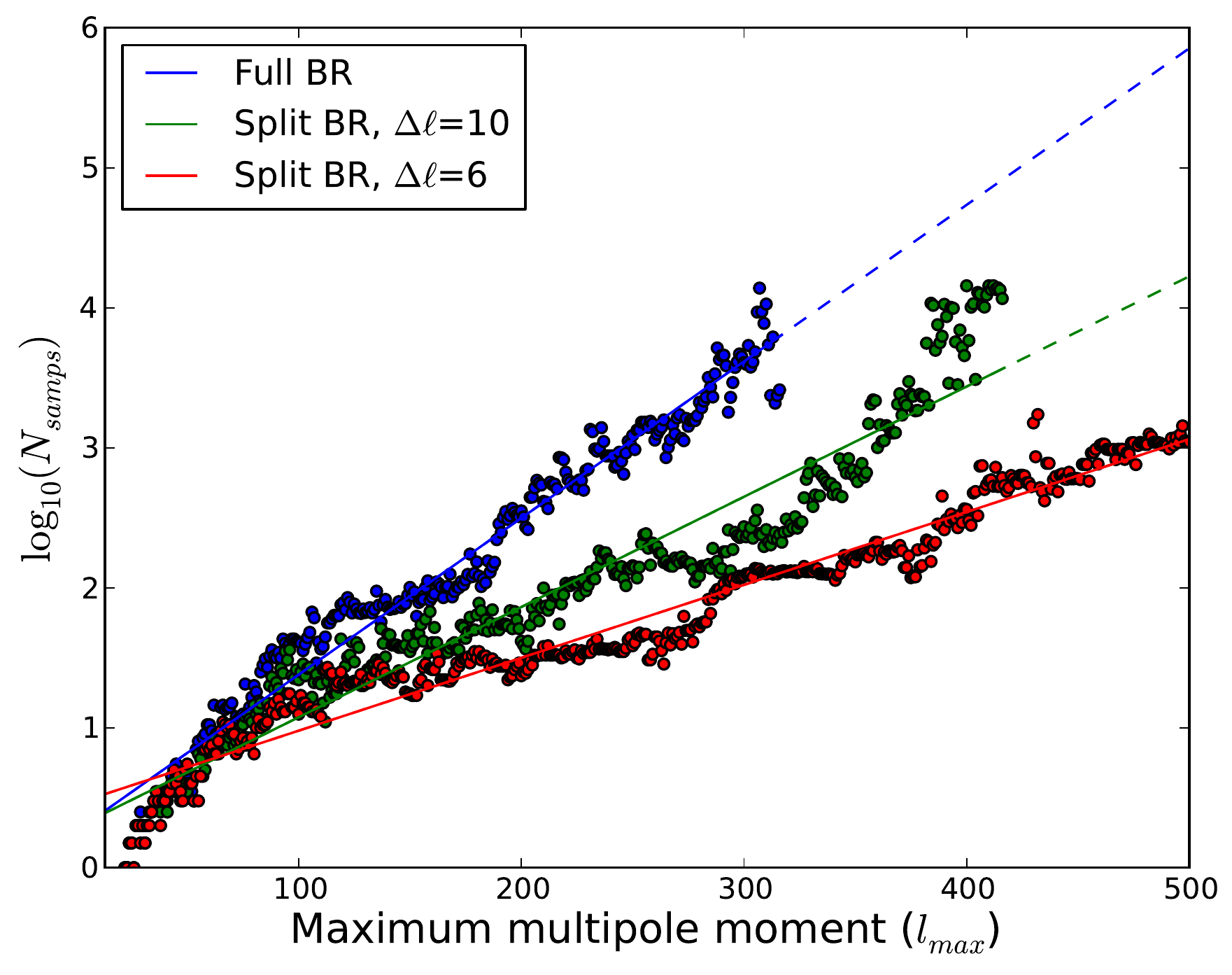}
    \includegraphics[width=\linewidth]{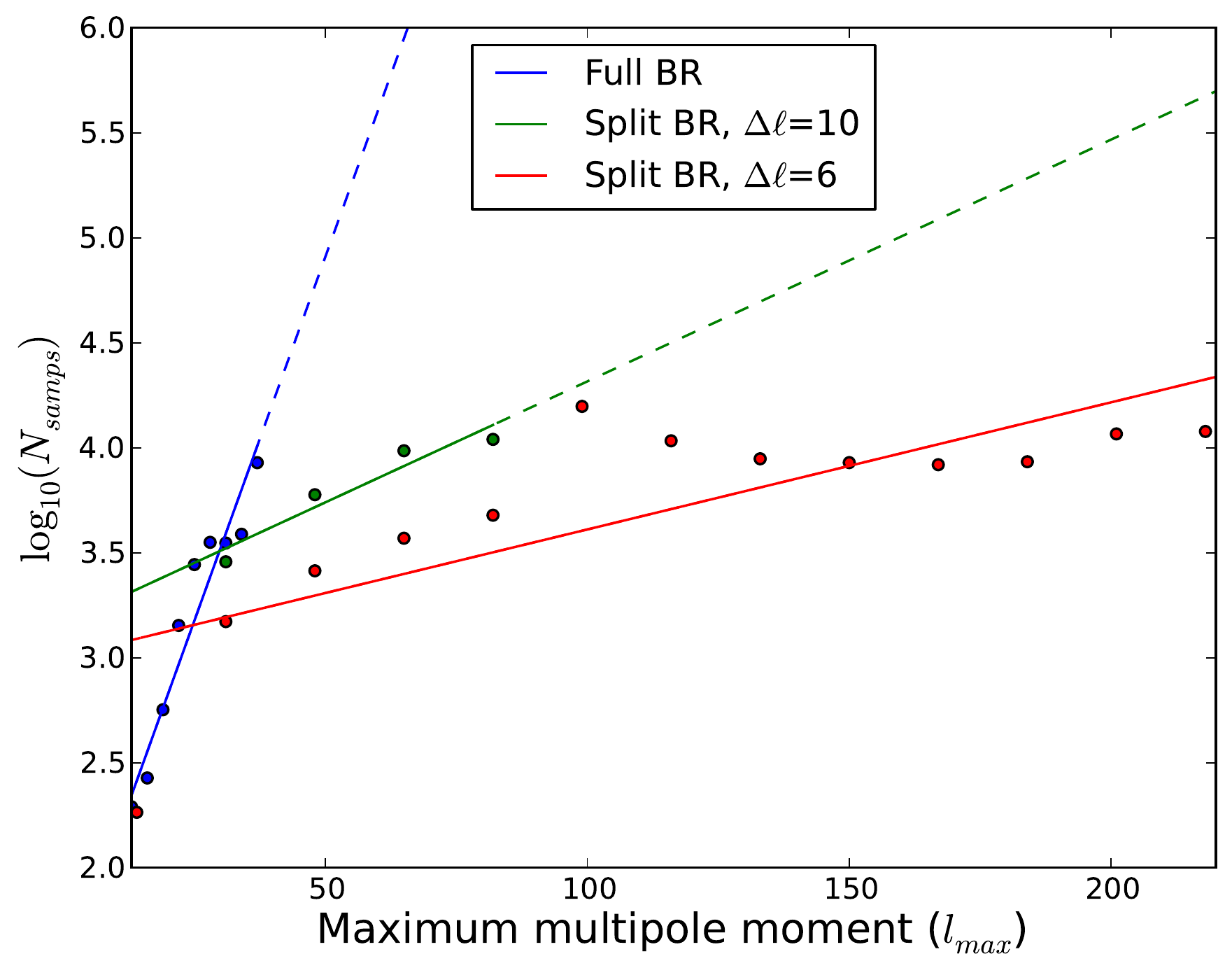}
  \end{center}
  \caption{Convergence analysis for the split Blackwell-Rao estimator, with convergence defined in sec. \ref{sec:bran}. The samples come from running Commander on a full-sky simulation. We show the median of the number of samples needed for convergence for a given $\ell_{\mathrm{max}}$, along with the best-fit regression line in $\log_{10}$-space. The median is computed from 10 (top) and 1024 
  (bottom) runs where the samples are scrambled between each run. The regression lines are dotted when they extend past the available data points. 
  The high number of runs per data point for the bottom plot is also the
  reason for the more sparse sampling - each data point represented a very
  high computational cost, and so the number of data points were reduced.}
  \label{fig:convergence_hires_fullsky}
\end{figure}

The block factorized BR estimators remain valid to higher
$\ell_{\textrm{max}}$, demonstrating how the ``curse of
dimensionality'' is lifted by breaking the full parameter space into
smaller regions that are easier to handle. In particular, the case
with $\Delta\ell=1$ agrees with the analytic case even at
$\ell_{\textrm{max}} = 900$ to $\sim$$0.3\sigma$.

In Figure~\ref{fig:maskedpoc} we show similar results for the
low-resolution simulation for which 25\,\% of the sky is removed by
masking, but this time comparing with the brute-force pixel-based
likelihood estimator, and this time using $N_{\textrm{samp}} = 60,000$ samples.
Again, we see that all cases agree to better
than $0.1\sigma$, even for the factorized BR estimator with
$\Delta\ell=5$, demonstrating the accuracy of both the full and the
factorized BR estimators, even with very small block sizes and for the
fairly large \emph{WMAP} mask.

Next, in the top panel of Figure~\ref{fig:convergence_hires_fullsky}
we plot the number of samples required for convergence according to
the above criterion for the high-resolution full-sky simulation
described above, and in the bottom panel we show the same, but after
applying the \emph{WMAP} mask, in order to introduce a realistic
multipole correlation structure. The upper vertical limit in these
plots is set by the finite number of samples included in the analysis.

In all cases we see the same qualitative behaviour: Reducing the
dimensionality of the BR estimator through block factorization greatly
improves the convergence rate by reducing the required number of
samples by orders of magnitude at high $\ell$'s. For instance, for the
full-sky case and with a block size of $\Delta\ell=6$, only $10^3$
samples are required in order to reach convergence up to
$\ell_{\textrm{max}}=500$, whereas the full BR estimator would require
$10^6$. For the 25\,\% \emph{WMAP} mask, about $10^4$ samples are
required for $\ell_{\textrm{max}}=200$, while it is difficult to
establish any sensible estimate for the full BR estimator in this
case. (Note that the high-$\ell$ projection for the latter case,
marked by a dashed line, is based on linear extrapolation from a few
low-$\ell$ points, since convergence was not reached at all within the
current sample set at higher multipoles. This projection is therefore
associated with a very large systematic uncertainty.)

\section{Conclusions}
\label{sec:discussion}

The main result presented in this paper is a statistically well
motivated block factorization of the CMB power spectrum
likelihood. Because the spherical harmonics are nearly orthogonal over
the large sky coverages achieved by current CMB satellite experiments
such as \emph{Planck} and \emph{WMAP}, any correlations between
different $C_{\ell}$s are localized in multipole space. Under the
assumption that these probabilistic dependencies have a strictly
finite range, the full CMB likelihood may be reduced into a product of
lower-dimensional marginals.

We have applied this result to two outstanding problems in CMB
analysis. First, we use this expression to derive a well-motivated
hybrid CMB likelihood estimator, merging an exact low-$\ell$ component
with an approximate high-$\ell$ component, that accounts for
correlations between the two regions. Although a detailed analysis of
the \emph{WMAP} likelihood shows that these correlations are
negligible for the \emph{WMAP} sky cut and the six-parameter
$\Lambda$CDM model, we nevertheless recommend this new estimator for
future experiments and analyses, both because its implementation is
trivial, and because it provides additional safety when analyzing
non-standard models. 

Second, we have shown how the same expression may be used to
accelerate the convergence rate of the Blackwell-Rao CMB likelihood
estimator by orders of magnitude at high $\ell$s. This is achieved by
factorizing the full parameter space into subspaces that each
individually converge faster, and then merging these sub-blocks into a
full-range estimator at the likelihood level using the block
factorization formula.

It should be noted that these results rely directly on the assumption
of vanishing long-range correlations. While this assumption holds to a
very high accuracy for the basic CMB signal plus noise data model, it
is in general not valid when including systematic effects in the
analysis. Perhaps the two most important examples are correlated beam
uncertainties and unresolved extra-Galactic point sources, each of
which extend through all $\ell$'s
\citep[e.g.,][]{planck:2013_P08}. Fortunately, these long-range
degrees of freedom may be modelled in terms of a small number of power
spectrum templates, each with an unknown amplitude. One can therefore
marginalize over these by sampling the unknown amplitudes as nuiscance
parameters, similar to what was done for high-$\ell$ astrophysical
parameters in the 2013 \emph{Planck} likelihood
\citep{planck:2013_P08}.

Finally, we note that the block factorization presented in
Section~\ref{sec:factorization} is a completely general statistical
result that holds exactly for any banded probability distribution, and
we therefore expect it to also find applications outside the CMB
field. 

\begin{acknowledgements}
This project was supported by the ERC Starting Grant StG2010-257080.
Part of the research was carried out at the Jet Propulsion Laboratory,
California Institute of Technology, under a contract with NASA. Some
of the results in this paper have been derived using the HEALPix
\citep{gorski:2005} software and analysis package.
\end{acknowledgements}

\end{document}